\PassOptionsToPackage{dvipdfmx}{graphicx}
\PassOptionsToPackage{dvipdfmx}{color}
\documentclass{jps-cp}

\usepackage{txfonts}
\usepackage{bm}
\usepackage{braket}
\usepackage{newtxmath}
\usepackage{graphicx}
\usepackage{color}

\title{Variational Monte Carlo Study of Superconductivity in a Correlated System with Broken Inversion Symmetry}

\author{Tsuyoshi \textsc{Imazu} and Katsunori \textsc{Kubo}}

\inst{Advanced Science Research Center, Japan Atomic Energy Agency, Tokai, Ibaraki 319-1195, Japan}

\email{imazu.tsuyoshi@jaea.go.jp}

\recdate{September 20, 2026}

\abst{Unconventional superconductivity in systems with broken inversion symmetry has attracted considerable interest because antisymmetric spin-orbit coupling (ASOC) can strongly affect the spin structure and stability of superconducting states. In this study, we investigate superconductivity in a correlated system with the symmetry of a honeycomb network using the variational Monte Carlo method. We consider a single-band Hubbard model including an on-site Coulomb interaction $U$ and ASOC with coupling strength $\lambda$, and compare several pairing states with different pairing symmetries. The trial wave functions are constructed from particle-number- and Gutzwiller-projected BCS states. To examine the energetic stability of each pairing state, we calculate the variational energy as a function of the gap amplitude, while optimizing the variational parameter $\mu$ and the
Gutzwiller parameter $g$ for each gap amplitude. At $U/t = 10$, the $f$-wave state has the lowest variational energy for $\lambda/t = 0$, whereas the chiral $p$-wave state becomes the most energetically favorable for $\lambda/t = 0.1$. These results indicate that ASOC strongly affects the competition among superconducting pairing states in the presence of strong electron correlations.}

\kword{superconductivity, chiral superconductivity, antisymmetric spin-orbit coupling, variational Monte Carlo}

\begin{document}
\maketitle

\section{Introduction}

Unconventional superconductivity in correlated electron systems has
been one of the central topics in condensed-matter physics. In
particular, chiral superconducting states have attracted considerable
attention because they spontaneously break time-reversal symmetry and
can possess nontrivial topological properties. Representative examples
are the spin-triplet chiral $p$-wave state and the spin-singlet chiral $d$-wave state. Both states are characterized by complex
multicomponent order parameters, although they have different spin
and momentum structures \cite{Sigrist1991,Read2000,BlackSchaffer2014}.

Systems with trigonal or hexagonal crystal symmetry provide a natural
platform for such chiral superconducting states, since multidimensional
irreducible representations allow degenerate pairing components and
their chiral combinations. Electron correlations can stabilize
unconventional pairing states in these systems, and several competing
pairing channels can become energetically close. For example, chiral
$d$-wave superconductivity has been extensively discussed in correlated
honeycomb and triangular-lattice systems
\cite{BlackSchaffer2014,BlackSchaffer2014PRB}.
Spin-triplet pairing, including $p$-wave states, is also of particular
interest in correlated systems because it provides a route to
topological superconductivity.

The absence of spatial inversion symmetry provides another important
ingredient in determining the superconducting state. In
noncentrosymmetric systems, antisymmetric spin-orbit coupling (ASOC) is
allowed by symmetry and lifts the spin degeneracy of the electronic
bands at generic momenta \cite{Smidman2017}. Since chiral $p$- and
chiral $d$-wave states have different spin and momentum structures,
their energetic responses to ASOC need not be the same. Therefore,
when strong electron correlations and ASOC coexist, their interplay
can substantially modify the competition among different
superconducting pairing states.
The noncentrosymmetric superconductor
$\mathrm{BaPtAs}_{1-x}\mathrm{Sb}_x$, which contains an ordered
honeycomb network of Pt and pnictogen atoms, provides a motivation for
studying such competition \cite{Kudo2018Sb,Kudo2018As,Ogawa2022}. Muon-spin-relaxation measurements on BaPtSb
have revealed a spontaneous internal magnetic field below the
superconducting transition temperature, suggesting
time-reversal-symmetry-breaking superconductivity
\cite{Adachi2025}. Previous theoretical studies have also shown that
several unconventional pairing states can compete in this material
family \cite{Imazu2026}.

Motivated by these studies, we investigate superconductivity in a
correlated system with the symmetry of the honeycomb network using the
variational Monte Carlo method \cite{Yokoyama1987,Gros1988}. To construct a minimal model that retains the relevant lattice symmetry, we consider an effective triangular-lattice model with an on-site Coulomb interaction and ASOC. We compare extended $s$-wave,
chiral $d$-wave, $f$-wave, and chiral $p$-wave pairing states
and examine their variational energies. Our aim is to clarify how ASOC affects the energetic competition among different pairing states 
in the presence of strong electron correlations.

\section{Model and Method}

\begin{figure}[t]
    \centering
    \includegraphics[width=0.8\linewidth]{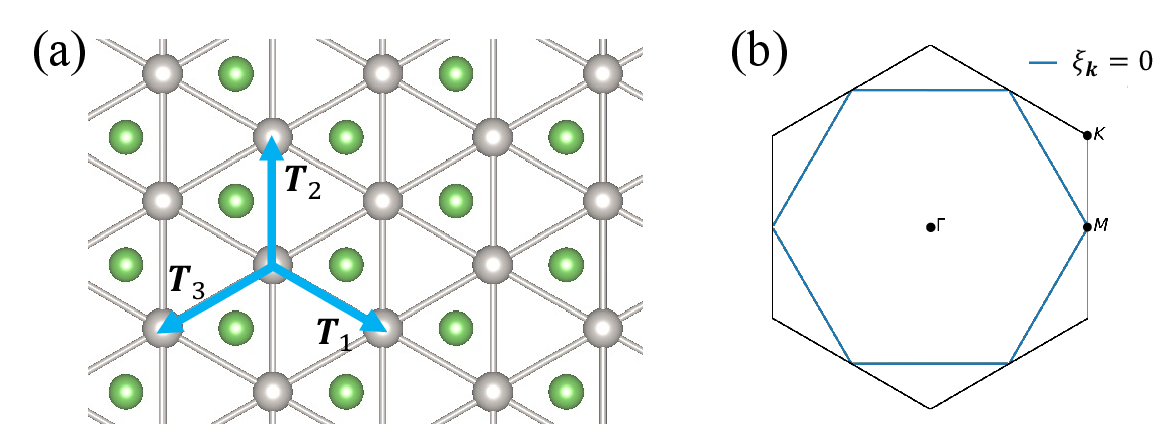}
    \caption{
    (a) Triangular-lattice model considered in the present study.
    The vectors $\bm{T}_i$ ($i=1,2,3$) denote the three
    nearest-neighbor directions.
    (b) Fermi surface of the noninteracting system in the
    thermodynamic limit at quarter filling for $\lambda/t=0$.
    }
    \label{fig:model}
\end{figure}

To investigate the interplay between electron correlations and
ASOC, we consider a minimal
single-band model on a triangular lattice that retains the relevant
symmetry of the honeycomb network.
The triangular lattice and the three nearest-neighbor vectors
$\bm{T}_i$ ($i=1,2,3$) are illustrated in Fig.~\ref{fig:model}(a).

The Hamiltonian is given by
\begin{align}
    &H=H_{0}+H_{\rm{ASOC}}+H_U \\
    &H_{0}=\sum_{\boldsymbol{r},\sigma} \sum_{i=1}^3 \Big[ t  \psi^{\dag}_{\sigma}(\boldsymbol{r}+\boldsymbol{T}_i)\psi_{\sigma}(\boldsymbol{r})+h.c  \Big]\\
&H_{\rm{ASOC}}=\sum_{\boldsymbol{r},\sigma}\sum_{i=1}^3 \Big[i \lambda \sigma\psi^{\dag}_{\sigma}(\boldsymbol{r}+\boldsymbol{T}_i)\psi_{\sigma}(\boldsymbol{r})+\mathrm{h.c.}     \Big] \\
&H_U=U\sum_{\boldsymbol{r}}n_{\uparrow}(\boldsymbol{r})n_{\downarrow}(\boldsymbol{r})
\end{align}
where $n_{\sigma}(\boldsymbol{r})= \psi_{\sigma}^{\dag}(\boldsymbol{r}) \psi_{\sigma}(\boldsymbol{r})$ is the electron density at site $\bm{r}$.
Here, $t$, $\lambda$, and $U$ denote the nearest-neighbor hopping,
the ASOC strength, and the on-site Coulomb interaction, respectively.
We take $t$ as the unit of energy and use $\sigma=+1$ ($-1$)
for spin up (down).

In momentum space, the ASOC is characterized by
\begin{align}
    \bm{\varg}_{\bm{k}}
    =
    2\sum_{i=1}^{3}
    \sin(\bm{k}\cdot\bm{T}_i)\hat{\bm{z}},
    \label{eq:gk}
\end{align}
which behaves around the $\Gamma$ point as
\begin{align}
    \varg_{\bm{k},z}
    \propto
    k_y(k_y^2-3k_x^2).
\end{align}
The corresponding single-particle dispersion used in the BCS
wave function is
\begin{align}
    \xi_{\bm{k},\sigma}
    =
    \varepsilon_{\bm{k}}
    -\mu
    +\lambda\sigma \varg_{\bm{k},z},
    \label{eq:dispersion}
\end{align}
where
\begin{align}
    \varepsilon_{\bm{k}}
    =
    2t\sum_{i=1}^{3}
    \cos(\bm{k}\cdot\bm{T}_i).
    \label{eq:bare_dispersion}
\end{align}
Here, $\mu$ is a variational parameter that controls the underlying
BCS wave function and is not included in the Hamiltonian used to
evaluate the variational energy.
For finite $\lambda$, the ASOC lifts the spin degeneracy at a given
momentum while preserving time-reversal symmetry.

In the present study, we focus on quarter filling,
$N_{\rm e}/(2N_{\rm s})=0.25$, where $N_{\rm e}$ and $N_{\rm s}$
denote the numbers of electrons and lattice sites, respectively.
The VMC calculations are performed on an
$L_1\times L_2=10\times10$ lattice with
$N_{\rm s}=100$ sites and $N_{\rm e}=50$ electrons.
Figure~\ref{fig:model}(b) shows the Fermi surface of the
noninteracting system in the thermodynamic limit at quarter filling
for $\lambda/t=0$.

We investigate the superconducting states using the VMC method.
The trial wave function is written as
\begin{align}
    |\Phi_{\rm VMC}\rangle
    =
    \mathcal{P}_{N}\mathcal{P}_{G}
    |\phi_{\rm BCS}\rangle ,
    \label{eq:vmc}
\end{align}
where $\mathcal{P}_{N}$ is the particle-number projection operator
that fixes the total number of electrons to $N_{\rm e}$, and
$\mathcal{P}_{G}$ is the Gutzwiller factor,
\begin{align}
    \mathcal{P}_{G}
    =
    \prod_{\bm{r}}
    \left[
    1-(1-g)
    n_{\uparrow}(\bm{r})n_{\downarrow}(\bm{r})
    \right].
    \label{eq:gutz}
\end{align}
The Gutzwiller factor suppresses doubly occupied configurations, and the variational parameter $g$ controls their weight \cite{Gutzwiller1965}.
In the present study, the filling is fixed at $N_{\rm e}/(2N_{\rm s})=0.25$
through the particle-number projection $\mathcal{P}_N$.

The BCS wave function is given by
\begin{align}
    |\phi_{\rm BCS}\rangle
    =
    \prod_{\bm{k}}
    \left(
    u_{\bm{k}}
    +
    v_{\bm{k}}
    c^{\dagger}_{\bm{k}\uparrow}
    c^{\dagger}_{-\bm{k}\downarrow}
    \right)
    |0\rangle ,
    \label{eq:bcs}
\end{align}
with
\begin{align}
    \frac{v_{\bm{k}}}{u_{\bm{k}}}
    =
    \frac{\Delta f_{\bm{k}}}
    {
    \xi_{\bm{k},\uparrow}
    +
    \sqrt{
    \xi_{\bm{k},\uparrow}^{2}
    +
    \Delta^{2}|f_{\bm{k}}|^{2}
    }
    }.
    \label{eq:vkuk}
\end{align}
Here,
$\Delta$ is a variational parameter in the underlying BCS wave function, and $f_{\bm{k}}$ describes the momentum dependence of the gap function.
In the present study, spin-singlet and spin-triplet pairing states
are treated separately, and singlet--triplet mixing is not included.
We consider four pairing states: extended $s$-wave, chiral $d$-wave,
$f$-wave, and chiral $p$-wave.
Their basis functions are summarized in Table~\ref{tab:gap}
\cite{Kudo2022,Imazu2026}.

\begin{table}[t]
\caption{
Basis functions $f_{\bm{k}}$ of the superconducting states considered
in the present study.
The last column shows their leading momentum dependence around the
$\Gamma$ point.
}
\label{tab:gap}
\centering
\begin{tabular}{c c c}
\hline\hline
State
& $f_{\bm{k}}$
& Leading form around $\Gamma$
\\
\hline
extended $s$
&
$\displaystyle
\sum_{n=1}^{3}
\cos(\bm{k}\cdot\bm{T}_n)
$
&
$\displaystyle
1- \frac{1}{4}(k_x^2+k_y^2)
$
\\[2ex]

chiral $d$
&
$\displaystyle
\sum_{n=1}^{3}
e^{-i2n\pi/3}
\cos(\bm{k}\cdot\bm{T}_n)
$
&
$\displaystyle
(k_x^2-k_y^2)+2ik_xk_y
$
\\[2ex]

$f$
&
$\displaystyle
\sum_{n=1}^{3}
\sin(\bm{k}\cdot\bm{T}_n)
$
&
$\displaystyle
k_y(k_y^2-3k_x^2)
$
\\[2ex]

chiral $p$
&
$\displaystyle
\sum_{n=1}^{3}
e^{i2n\pi/3}
\sin(\bm{k}\cdot\bm{T}_n)
$
&
$\displaystyle
k_x+ik_y
$
\\
\hline\hline
\end{tabular}
\end{table}

The variational energy is evaluated as
\begin{align}
    E(\mu,\Delta,g)
    =
    \frac{
    \langle\Phi_{\rm VMC}|H|\Phi_{\rm VMC}\rangle
    }{
    \langle\Phi_{\rm VMC}|\Phi_{\rm VMC}\rangle
    }.
    \label{eq:energy}
\end{align}
To examine the energetic stability of each pairing state,
we fix $\Delta$ at a series of values and optimize
$\mu$ and $g$ independently for each $\Delta$.
We therefore define
\begin{align}
    E_{\rm opt}(\Delta)
    =
    \min_{\mu,g}
    E(\mu,\Delta,g),
    \label{eq:eopt}
\end{align}
and compare the resulting energy curves among the different pairing
symmetries.
The expectation values are evaluated using $5\times10^{7}$
Monte Carlo samples, which are divided into 50 bins for estimating
the statistical errors.

\section{Results}

\begin{figure}[t]
    \centering
    \includegraphics[width=0.95\linewidth]{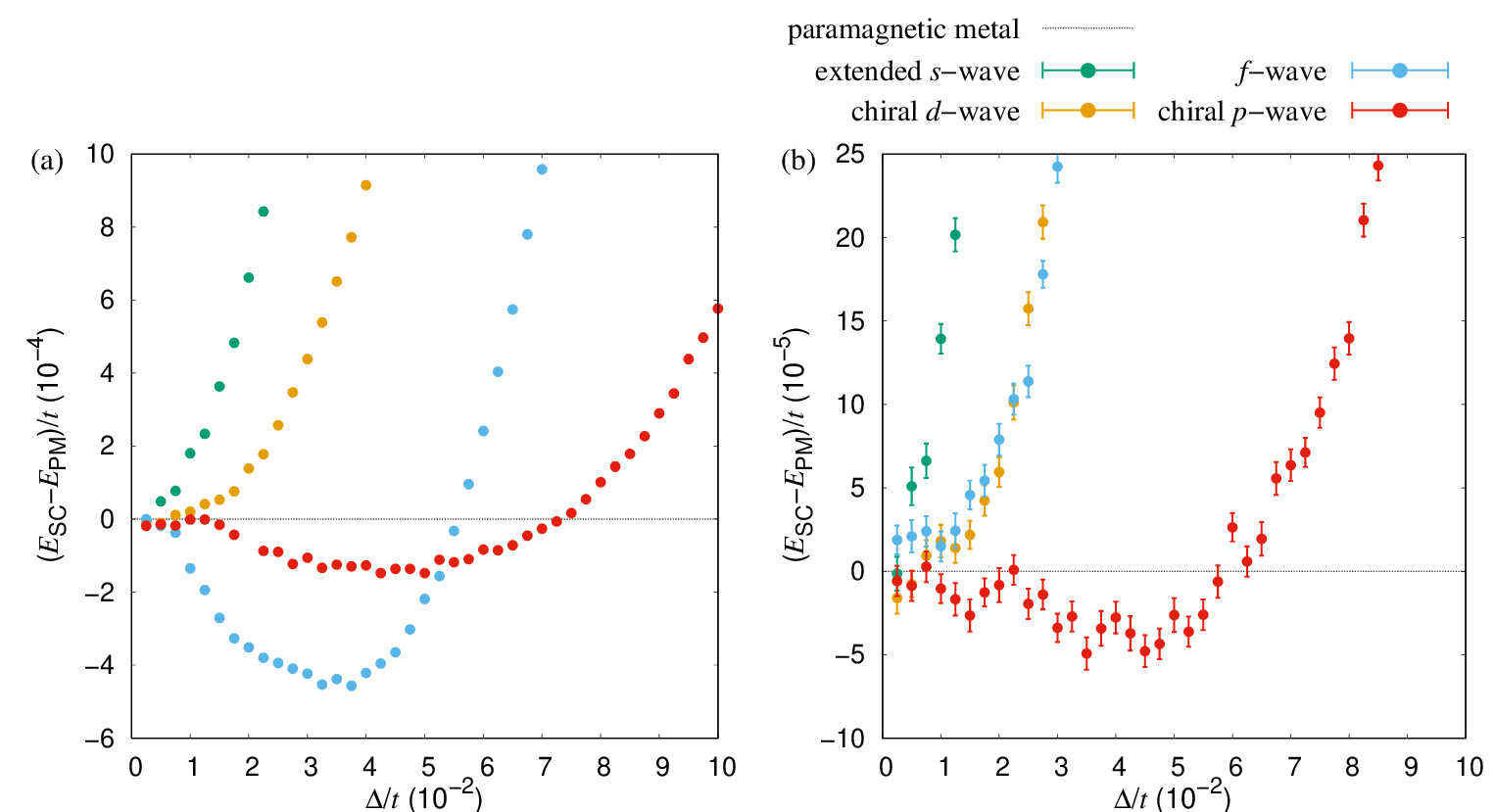}
    \caption{
Variational energy difference
$(E_{\rm SC}-E_{\rm PM})/t$ as a function of the pairing amplitude
$\Delta/t$ at $U/t=10$ and quarter filling for
(a) $\lambda/t=0$ and (b) $\lambda/t=0.1$.
Here, $E_{\mathrm{SC}}$ and $E_{\mathrm{PM}}$ denote the variational energies obtained from the projected BCS and projected Fermi-sea trial wave functions, respectively.
For each value of $\Delta$, the variational parameter $\mu$ and the Gutzwiller parameter $g$
are optimized independently.
The horizontal dashed line represents $E_{\rm SC}=E_{\rm PM}$.
The vertical scales are $10^{-4}$ in panel (a) and $10^{-5}$ in
panel (b).
Error bars represent the statistical errors estimated from the
Monte Carlo sampling.
    }
    \label{fig:energy}
\end{figure}

Figure~\ref{fig:energy} shows the variational energy difference
between the projected BCS and projected Fermi-sea states,
$(E_{\rm SC}-E_{\rm PM})/t$, as a function of the pairing amplitude
$\Delta/t$ at $U/t=10$ and quarter filling.
Here, $E_{\mathrm{SC}}$ and $E_{\mathrm{PM}}$ denote the variational energies obtained from the projected BCS and projected Fermi-sea trial wave functions, respectively. 
For each fixed value of $\Delta$, the variational parameter
$\mu$ and the Gutzwiller parameter $g$ are optimized independently.
Thus, a negative value of $(E_{\mathrm{SC}}-E_{\mathrm{PM}})/t$ indicates that the corresponding pairing state is energetically favored over the projected Fermi-sea state. 
Note that the vertical scales are different between the two panels:
the energy difference is shown on the scale of $10^{-4}t$ in
Fig.~\ref{fig:energy}(a) and $10^{-5}t$ in
Fig.~\ref{fig:energy}(b).

In the absence of ASOC, $\lambda/t=0$, the four pairing states
exhibit qualitatively different dependences on $\Delta$.
The extended $s$-wave and chiral $d$-wave states rapidly increase
in energy with increasing $\Delta$ and do not show a sizable
energy gain relative to the paramagnetic state.
In contrast, the $f$-wave state exhibits a pronounced minimum at
a finite pairing amplitude,
$\Delta/t \simeq 0.035$--$0.04$.
The corresponding energy gain is approximately
$4.5\times10^{-4}t$.
The chiral $p$-wave state also has a finite range of $\Delta$
in which its energy is lower than $E_{\rm PM}$, with a broad
minimum of approximately $1.5\times10^{-4}t$.
Its energy gain is, however, considerably smaller than that of
the $f$-wave state.
Therefore, among the four trial states considered here, the
spin-triplet $f$-wave state is the most favorable state for
$\lambda/t=0$.
It is also noteworthy that both triplet pairing states considered
here, the $f$-wave and chiral $p$-wave states, gain energy relative
to the paramagnetic metallic state, whereas the singlet extended
$s$-wave and chiral $d$-wave states do not show a comparable
stabilization. This tendency is consistent with a previous VMC study
of the quarter-filled triangular-lattice Hubbard model by Kataoka and
Ohta, in which a spin-triplet state was found to have a lower energy
than both spin-singlet states and the normal state
\cite{Kataoka2008}.

The competition among the pairing states changes substantially
when a finite ASOC is introduced.
At $\lambda/t=0.1$, the $f$-wave state no longer exhibits the
large energy gain found at $\lambda/t=0$.
Instead, the chiral $p$-wave state develops a clear minimum at
a finite pairing amplitude around
$\Delta/t \simeq 0.04$--$0.05$.
The minimum energy is approximately
$5\times10^{-5}t$ below that of the paramagnetic metal.
The extended $s$-, chiral $d$-, and $f$-wave states do not show
a comparable finite-$\Delta$ energy gain in this parameter region.
Consequently, the chiral $p$-wave state becomes the most favorable
among the superconducting trial states considered at
$\lambda/t=0.1$.

These results show that even a relatively small finite ASOC
significantly reorganizes the energetic competition among the
superconducting pairing channels.
In particular, the lowest-energy state changes from the $f$-wave
state at $\lambda/t=0$ to the chiral $p$-wave state at
$\lambda/t=0.1$.
It is also noteworthy that the characteristic
variational energy gain at $\lambda/t=0.1$ is approximately one order of
magnitude smaller than that of the $f$-wave state at
$\lambda/t=0$.
Thus, ASOC affects not only which pairing symmetry is favored but
also the energy scale associated with the stabilization of the  pairing state.
A complementary analysis of long-distance pairing correlations and their system-size dependence would further clarify whether these energetically favored pairing states exhibit superconducting long-range order\cite{Yokoyama2013,Watanabe2025}.

\section{Summary}

We investigated superconductivity in a correlated system with the
symmetry of a honeycomb network using the VMC
method. A minimal single-band triangular-lattice model with an
on-site Coulomb interaction and ASOC was considered, and the variational energies of extended
$s$-wave, chiral $d$-wave, $f$-wave, and chiral $p$-wave states
were compared at quarter filling and $U/t=10$. In the absence of
ASOC, $\lambda/t=0$, the two spin-triplet states gain energy relative
to the paramagnetic metallic state, whereas the spin-singlet states
do not show a comparable stabilization, and the $f$-wave state has
the lowest variational energy among the states considered. This
tendency toward triplet pairing is consistent with a previous VMC
study of the quarter-filled triangular-lattice Hubbard model
\cite{Kataoka2008}. 
For finite ASOC, $\lambda/t=0.1$, the energetic
competition is substantially modified and the chiral $p$-wave state
becomes
the most energetically favorable among the pairing states considered. 
These results
demonstrate that ASOC can reorganize the competition among
unconventional pairing states in the presence of strong electron
correlations.

\section*{Acknowledgment}

\begingroup
\sloppy
This work was partially supported by JSPS KAKENHI
(Grant Numbers JP26K24528 and JP23K03330).
Part of the numerical calculations was performed on the supercomputer system of the Japan Atomic Energy Agency.
\par
\endgroup

\end{document}